# Population II Distance Indicators

# RR Lyrae Variables
# Tip of the Red Giant Branch (TRGB) Stars
# and
# J-Branch Asymptotic Giant Branch (JAGB/Carbon) Stars

Barry F. Madore [1,2] and Wendy L. Freedman [2]

[1] The Observatories,
Carnegie Institution for Science, 813 Santa Barbara Street
Pasadena CA 91101 USA
email: barry.f.madore@gmail.edu

[2] Department of Astronomy and Astrophysics,
University of Chicago
5640 South Ellis Avenue
Chicago, IL 60637 USA
email: wfreedman@uchicago.edu

**Abstract**
We review the theoretical underpinnings, evolutionary status, calibrations and current applications of three bright Population II extragalactic distance indicators: Tip of the Red Giant Branch (TRGB) stars, RR Lyrae variables and J-Branch Asymptotic Giant Branch (JAGB/Carbon) stars. For $M_I$ (TRGB) = -4.05 mag the Hubble constant is determined to be

$$H_o = 69.8 +/- 0.6 \text{ (stat)} +/- 1.6 \text{ (sys) km/s/Mpc}.$$

# Introduction

Three distinct, late-time, evolutionary phases of low-mass (1.0 M☉) and intermediate-mass (1.5 to 4.0 M☉) stars have identifiable features in their luminosity distributions that allow them to be found and used as galactic and extragalactic distance indicators. From a stellar evolutionary point of view the stars making their way up the Red Giant Branch (RGB) reach a maximum luminosity that is predicted and well understood from first principle stellar astrophysics. The discontinuity in the RGB luminosity function marks the Tip of the Red Giant Branch (TRGB) after which stars rapidly fade in luminosity, landing on the ``Horizontal" Branch in which a color-selected subset of these stars are unstable to radial pulsations. These are the Population II RR Lyrae variables. Finally, Asymptotic Giant Branch (AGB) stars, having masses in the range of 1.4 to 4M☉, become extremely red (JAGB/Carbon) stars, found at a narrowly confined range of luminosities, brighter than either of the other two phases just discussed. These three distance indicators are shown in relationship to each other in the schematic color-magnitude diagram shown above in Figure 1.

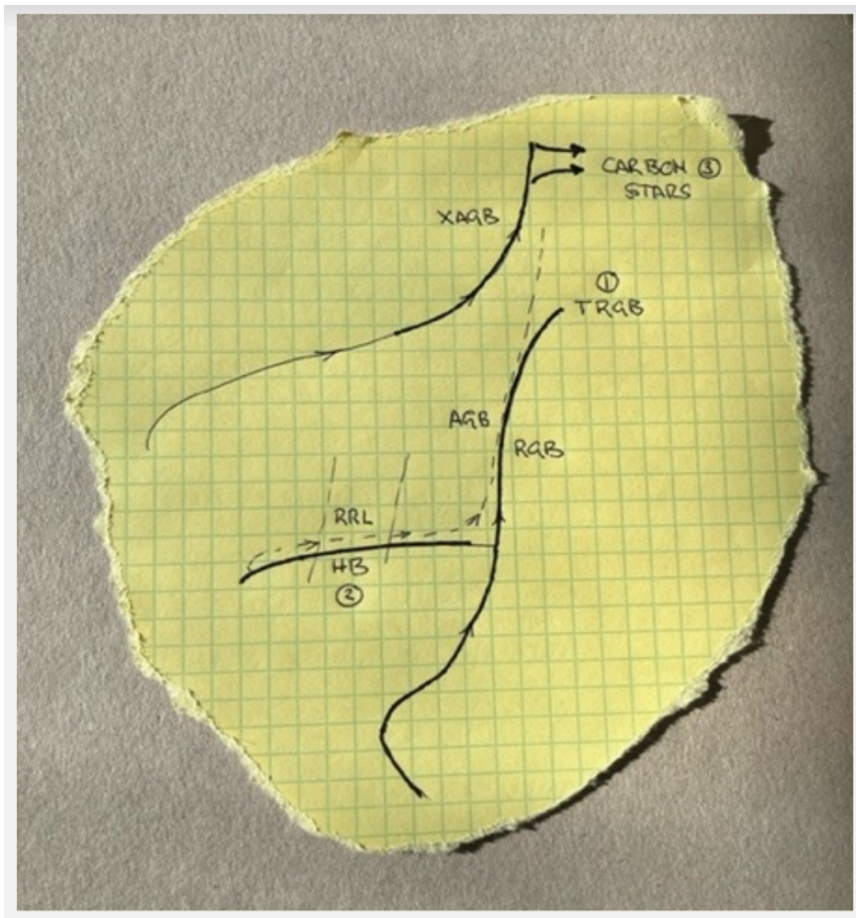

**Figure 1** A graphical representation of the relative disposition of the three stellar distance indicators discussed in this article.} (1) is the TRGB, found at intermediate luminosities and relatively red colors, marking the terminal point in the upward evolution of red giant branch stars. (2) is the so-called "Horizontal" Branch which crosses the vertical Population II variable star instability strip in which RR~Lyrae stars are found, and (3) is the domain of those highly evolved, extended asymptotic giant branch (X-AGB) stars, in the early carbon star phase, seen at the very highest luminosities and reddest colors.

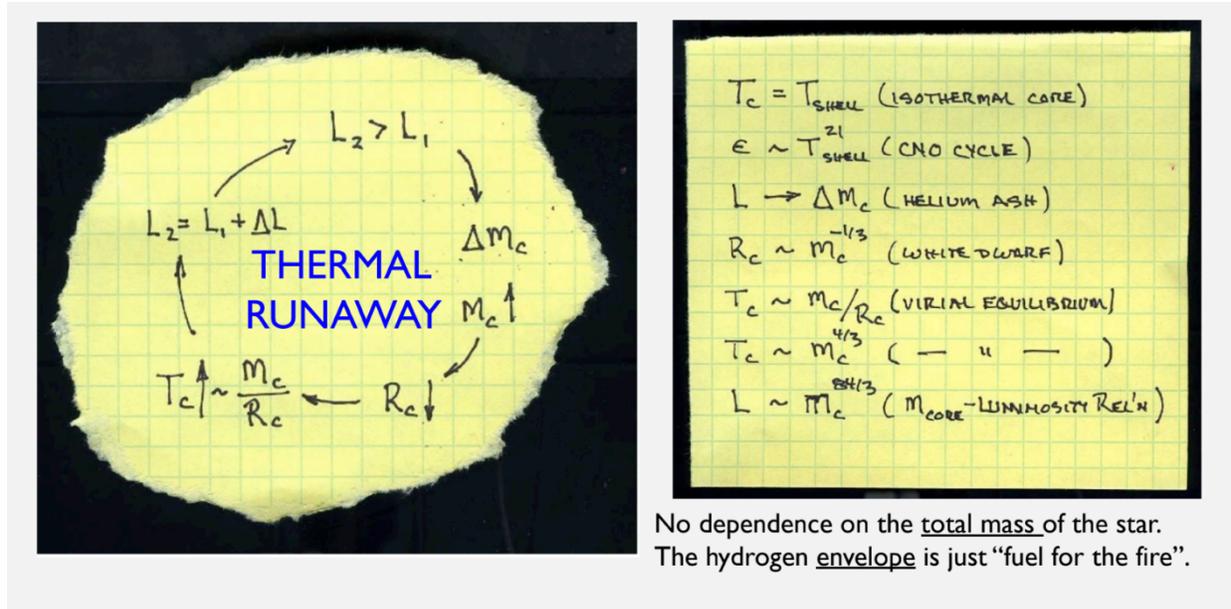

**Figure 2** Basics of the Evolution of Stars Ascending the Red Giant Branch: Left Panel: A schematic of the autocatalytic nuclear reaction leading to evolved low mass stars ascending the red giant branch. Right Panel: The inter-dependence of energy generation rates upon core mass, core radius, and shell temperature leading to a high sensitivity luminosity-core mass relation that is at the heart of the thermal runaway described in the left panel.

## Tip of the Red Giant Branch Stars: TRGB

TRGB stars are no strangers to crises in cosmology. Nearly a century ago the brightest red stars in globular clusters were being used both by Harlow Shapley and Heber Curtis, when they were constructing their absolute distance scales for the Milky Way. That was at a time when the Milky Way was considered to be the totality of our Universe, but its size was hotly debated (see Shapley & Curtis 1921, and especially Berendzen, Hart & Seeley 1986, Chapter 4 for a detailed discussion). Suffice it to say that their calibrations of the brightest (red giant) stars differed by 3 magnitudes! Within a few years Edwin Hubble (1925) proved both sides of that early crisis in cosmology to be wrong. Using Cepheids, discovered by him in *anagalactic nebulae,* (galaxies of comparable size to our Milky Way but external to it) he first showed the the Universe was far larger than anyone had anticipated, and then went on to demonstrate that this universe was expanding (Hubble 1929, 1936), at a rate of about 500 km/s/Mpc. Walter Baade (1948) subsequently used the brightest red giant stars to recalibrate the RR Lyrae zero point and the Cepheid distance scale to reduce the expansion rate of the Universe (the Hubble constant) by about a factor of two, thereby doubling the estimated age of the Universe.

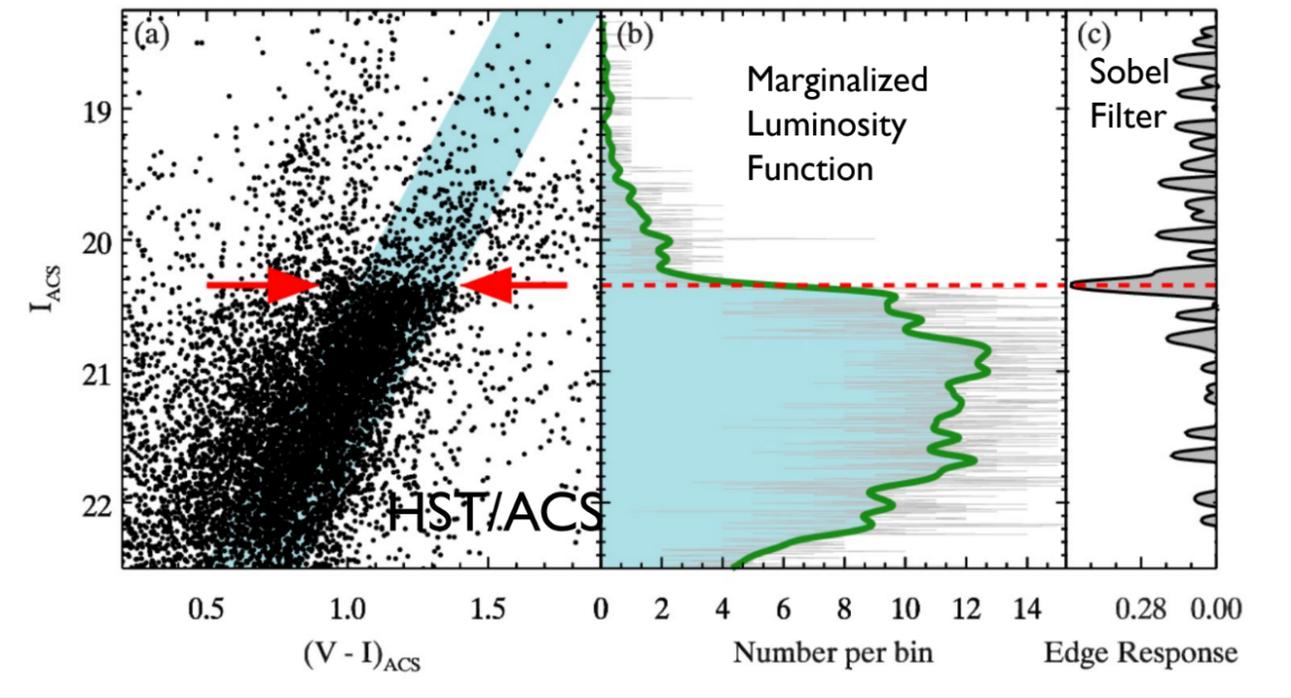

**Figure 3: The I-band Tip of the Red Giant Branch**. Left Panel (a): The I vs (V-I) CMD for the nearby galaxy IC 1613, showing the red giant branch terminating in a flat plateau at an I-band magnitude of about 20.3 mag. Middle Panel (b) shows the marginalized I-band luminosity function for the RGB stars found within the slanting blue shaded region in the CMD. The rapid rise in the luminosity function is obvious to the eye in both panels. Panel (c) shows the output of a digital (Sobel) filter that measures the first derivative of the luminosity function, with maximum response marking the TRGB.

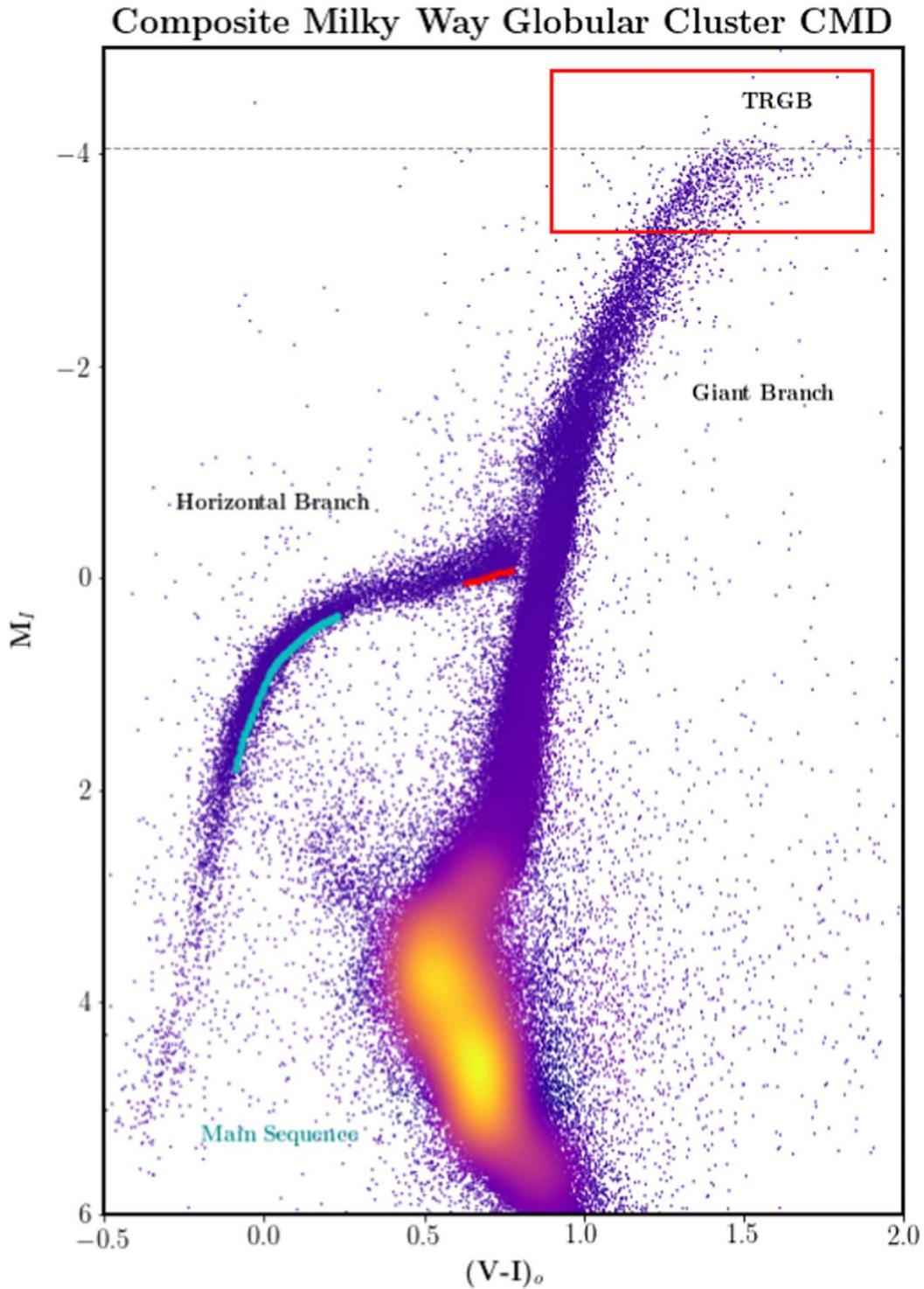

**Figure 4:** Composite I vs (V-I) CMD based on 46 Galactic Globular Clusters. The clusters span a range in metallicity of , -2.4 dex < [Fe/H] < -1.0 dex. Cluster membership was determined from their Gaia DR2 proper motions. The red rectangular box outlines the region of the TRGB. The horizontal dashed gray line indicates the TRGB at M_I = -4.056 mag. Adapted from Freedman (2021).

Today TRGB stars are playing a central role in the determination of distances to a wide variety of galaxies in the Local Volume far outstripping Cepheid distances in number\footnote{Close to 1,000 TRGB distances to about 300 galaxies, are compiled in NED, while less than 70 galaxies have Cepheid distances to date.} because of the simplicity of the TRGB Method in its application and its interpretation (see Figure 2): unlike Cepheids which are variable stars needing multi-epoch observations to determine periods, amplitudes and light curves before deriving mean magnitudes and colors (and the period-luminosity relations), the TRGB stars are non-variable, easily identified and have constant I-band magnitudes as a function of color/metallicity (see Figure 3). Figure 4 shows the absolute calibration, based on galactic globular clusters covering a wide range of metallicities (Cerny et al. 2020). TRGB stars are de facto *standard candles*; Cepheids and RR Lyrae variables (discussed next) are *standardizable candles*.

Most recently Freedman (2021) has reviewed the calibration of the zero point of the I-band TRGB method and its application to the extragalactic distance scale resulting in an independent value of the Hubble constant; see Figure 12.

## RR Lyrae Variables and the Horizontal Branch

RR Lyrae variables have been known for over a century and they have been used as standard candles almost from the beginning. However, because their evolutionary status places them where the (primarily vertical) Population II instability strip crosses the Horizontal Branch, their periods are largely irrelevant when it comes to predicting the absolute V-band magnitudes, where the Horizontal Branch is indeed flat/horizontal in the V-band CMD. (Note especially Figures 1.2, 1.7 and 3.2 in Horace Smith's (1995) comprehensive monograph entitled RR *Lyrae Stars*). That has led to the commonly held view that RR~Lyrae variables do not follow a Period-Luminosity relation at all, but the discussion devolved into one in which scatter in the absolute magnitude was seen to be a function only of the metallicity of the star (see Chapter 2 of Smith 1995). It therefore came as something of a surprise when Longmore, Fernley & Jameson (1986) explicitly followed the lead of McGonegal et al. (1985) by moving into the near infrared, publishing K-band PL relations for RR~Lyrae variables in globular clusters. These PL relations showed a steep dependence on period and remarkably low scatter. So the question arises: How is it possible that RR Lyrae variables are *standard candles* at one wavelength (the V band) and go on to becoming *standardizable candles,* being strongly dependent upon period at a longer (K-band) wavelength? The answer is to be found in the next section.

## Astrophysical (PLC) Data Cubes

The Stefan-Boltzmann relation applied to a spherical self-luminous body relates the total (bolometric) luminosity to only two parameters: the surface area of the sphere multiplied by the surface brightness; the latter being controlled by fourth power of the effective temperature. By direct analogy the luminosity of variable stars can be shown to also be the function of only two (observable) properties: the period (which substitutes for the radius, above) and the color (which substitutes naturally for the temperature.) The result is a Period-Luminosity-Color relation that can be applied conceptually to RR Lyrae variables and Cepheids, alike (Madore & Freedman 1990). The PLC is a tilted plane (of infinite extent but zero thickness) with the three dimensions being period, luminosity and color.

In Figure 5 we show this *Astrophysical Data Cube* for the PLC. The equation defining the plane is given at the top of the figure:  $V = a \log P + b (B-V) + c$

However, this infinite plane is limited by two additional constraints imposed by the independent physics of the instability strip: A blue edge, defined by the depth of the He++ ionization zone driving the pulsation, and a red edge, controlled by the onset of convection terminating the pulsation at cooler temperatures. The three orthogonal projections of this externally constrained PLC relation are described in the caption.

The slope of the PL relation (back face of the data cube) is defined by the slopes (D-B and/or E-A) of the edges of the instability strip, but a fair sampling in both period and especially magnitude must be had observationally if the slope is to be measured/calibrated empirically. A line of constant magnitude (C-A in the CMD, left face of the cube) is by definition, a line of constant magnitude in the marginalized PL relation (C-A). The slope of C-A is clearly not the slope of the PL relation as it offers an atypical and unrepresentative sampling of the instability strip.

In Figure 6 we show simulations of Horizontal Branch RR Lyrae variables (Catelan, Pritzl & Smith 2020) within the instability strip, whose boundaries are marked by upward slanting solid black lines. In keeping with the discussion of the {\it Astrophysical Data Cube} the red arrow shows the ridge line of the instability strip emphasizing the true slope of the PL relation and its non-coincidence with the slope of the majority of stars populating the Horizontal Branch.

## Reconciling Bailey Diagrams, Oosterhoff Groups and Sturch's Law

Sturch's Law (1966) states that at minimum light all RR Lyrae variables reach a low-dispersion in color that is a shallow function of period. As the right panel in Figure 7 shows, this ``law'' demands that the amplitudes of RR Lyrae variables must be a function of color (and period) in order for their mean magnitudes to place them along the Horizontal Branch. Moreover, stars passing over and above the Horizontal branch en *route* to the AGB will be forced into a second sequence of amplitudes as a function of period. Those two branches are known as Type I and Type II Oosterhoff types, and they are shown (suggestively rotated) in the left panel of Figure 7. See the wonderfully informative article on *RR Lyrae Period-Amplitude Diagrams* by Smith, Catelan & Kuehn (2011). It can thus be seen that the Bailey Diagrams, Oosterhoff Groups and Sturch's Law can all be understood on the basis of very simple and related underlying astrophysics, seen in different projections.

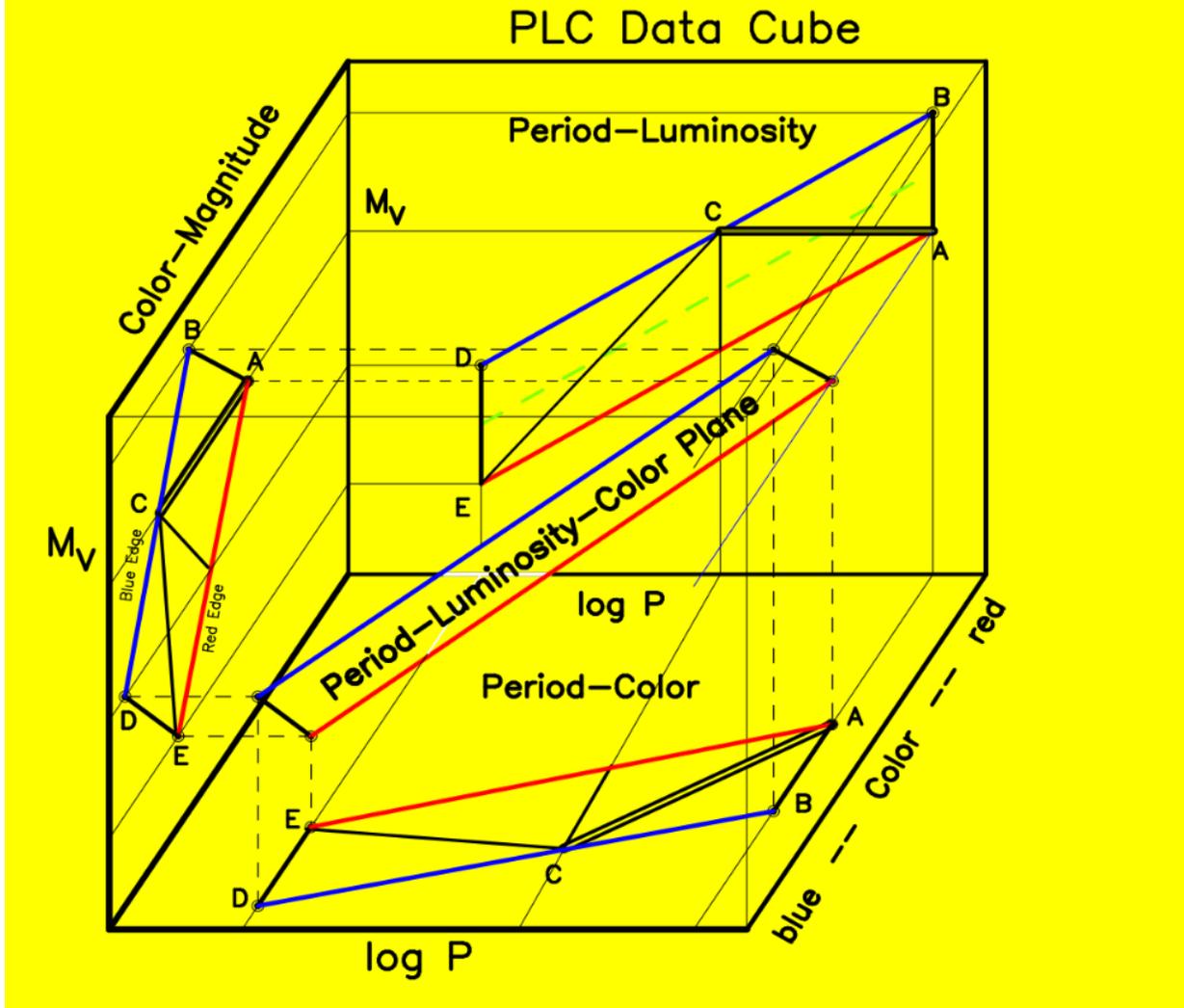

**Figure 5:** An Astrophysical Data Cube}, defined in three dimensions by the *Period-Luminosity-Color Relation.* Note that the PLC itself is a tilted plane in this 3-dimensional space. The instability strip is a (physically) constrained subset of the PLC plane, and various orthogonal marginalizations of that intersection form (a) the Period-Luminosity relation (face on), (b) the Period-Color relation (at the base) and also (c) the Cepheid instability strip within the Color-Magnitude Diagram (left face of the cube).

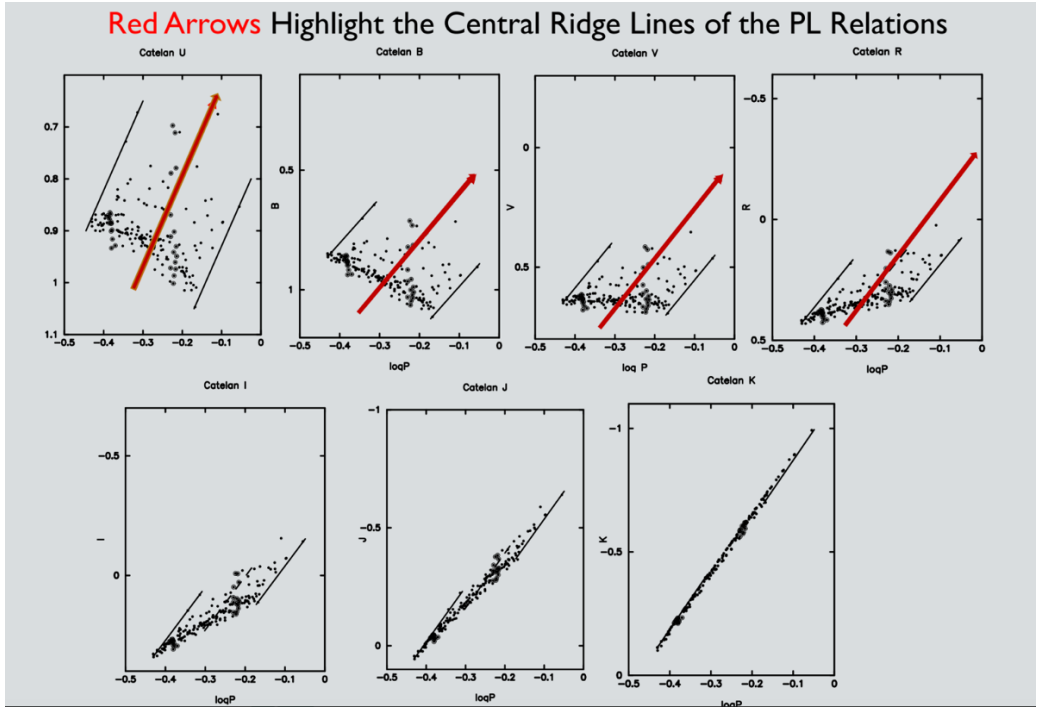

**Figure 6:** Multi-Wavelength period-luminosity relations for stars crossing the instability strip as derived from stellar evolution and pulsation models published by Catelan, Pritzl & Smith (2020).

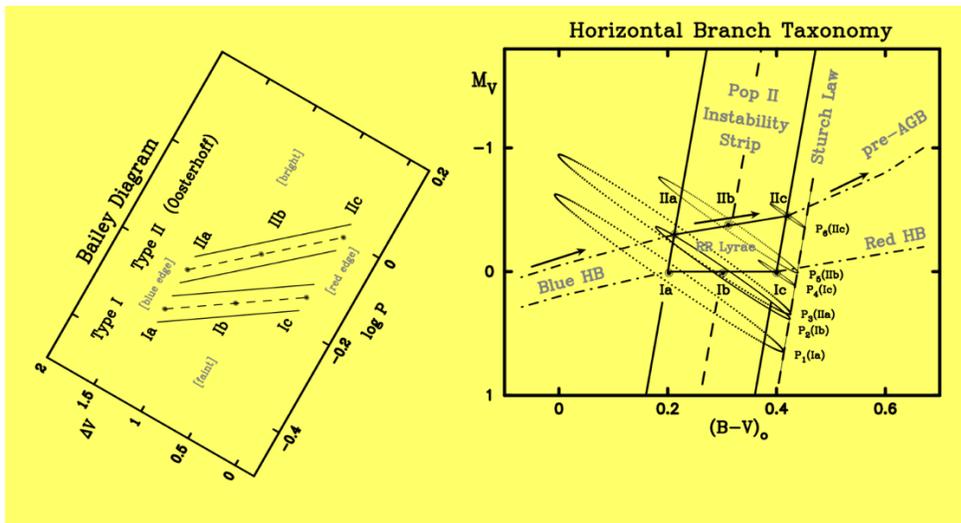

**Figure 7:** Unification of Bailey Diagrams, Oosterhoff Types, Sturch's Law and RR Lyrae Amplitudes within the Population II Instability Strip.} The Bailey Diagram (on the left) has been rotated counter-clockwise to draw attention to the fact that the two Oosterhoff sequences, of Type I and II, correspond to the Zero-Age Horizontal Branch (ZAHB) and the Super-Horizontal Branch stars (evolving across the instability strip from the blue to the red) above the ZAHB and into their AGB phase further to the red, as schematically shown in the CMD to the right.

# J Branch AGB (Carbon) Stars

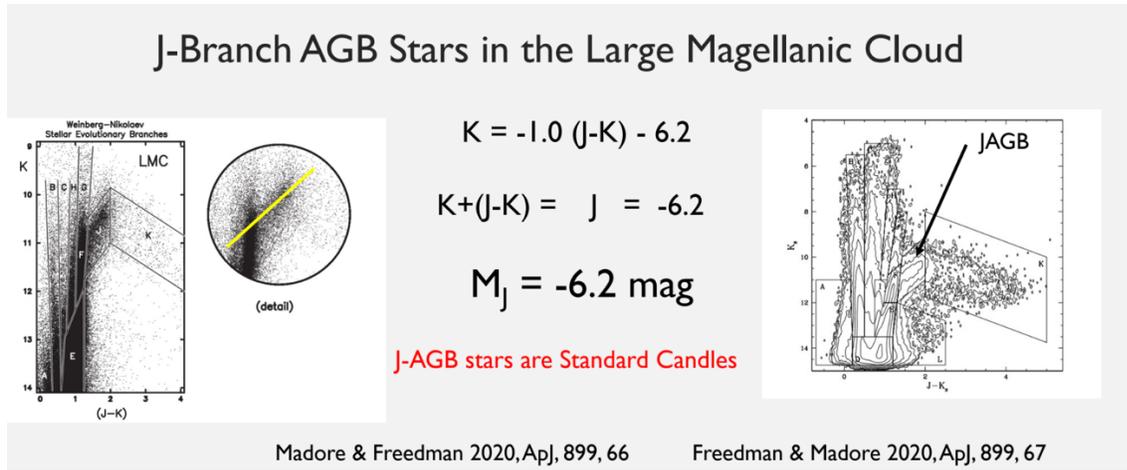

**Figure 8:** Discovery and Calibration of JAGB Stars in the Large Magellanic Cloud. Using three-band (JHK) near-infrared photometry from 2MASS Weinberg & Nikolaev (2001) were the first to classify the various sequences of stars in the K-(J-K) CMD and drew particular attention to the J Branch stars as potentially useful distance indicators, given the strong correlation of K-band magnitude with (J-K) color.

This final section discusses the most luminous intermediate-age stars in the asymptotic giant branch (AGB) phase: The J-Branch/Carbon stars first identified as high-luminosity, high-precision distance indicators by Nikolaev & Weinberg (2000). As the left section of Figure 8 illustrates, for stars in the LMC, the NIR K vs (J-K) CMD has a distinct feature where the K-band magnitude is seen to correlate tightly with (J-K) color with a slope of unity. As later pointed out by Madore & Freedman (2020) this means that the J-band magnitude is constant with color, making the JAGB stars *actual* standard candles (see equations in Fig. 8).

The very red color of the JAGB stars is caused by the introduction of freshly produced carbon into the outer envelopes and atmospheres of these already cool and extended stars. The contamination begins to reach the surface on the third (and later) thermal pulsing events. Dredge-up of carbon starts at a stellar mass of about 1.4 M☉ and would continue for all higher mass AGB stars were it not for the increased temperatures at the base of the convective zone that now burn the carbon before it can reach the surface. ``Hot-Bottom Burning'' commences at a stellar mass of about 4 M☉. These two constraints on the masses of AGB precursor stars that can become carbon stars set high and low luminosity limits which result in the JAGB stars becoming standard candles (see Figures 9 and 10).

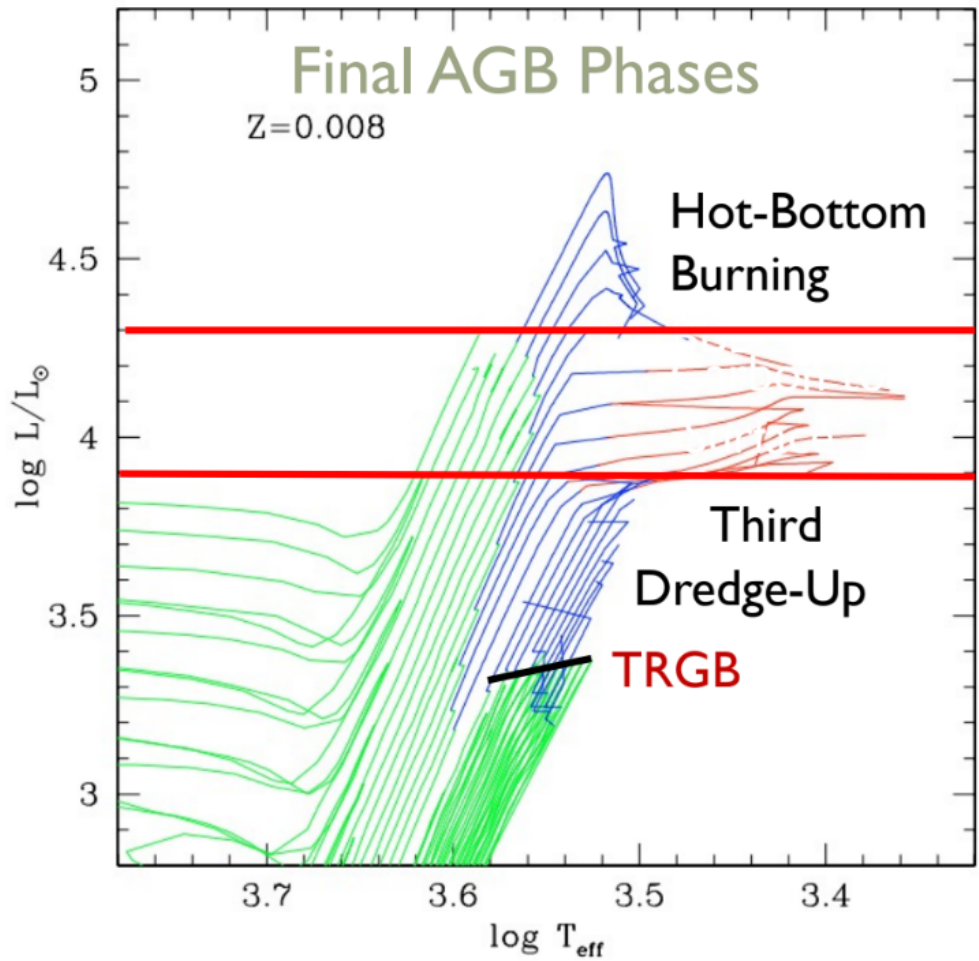

}

**Figure 9:** JAGB/Carbon Stars. Theoretical evolutionary tracks of intermediate-age, intermediate-mass Asymptotic Giant Branch stars entering the carbon star phase to the right at log $T_{eff}$ < 3.5. Adapted from Marigo et al. (2008).

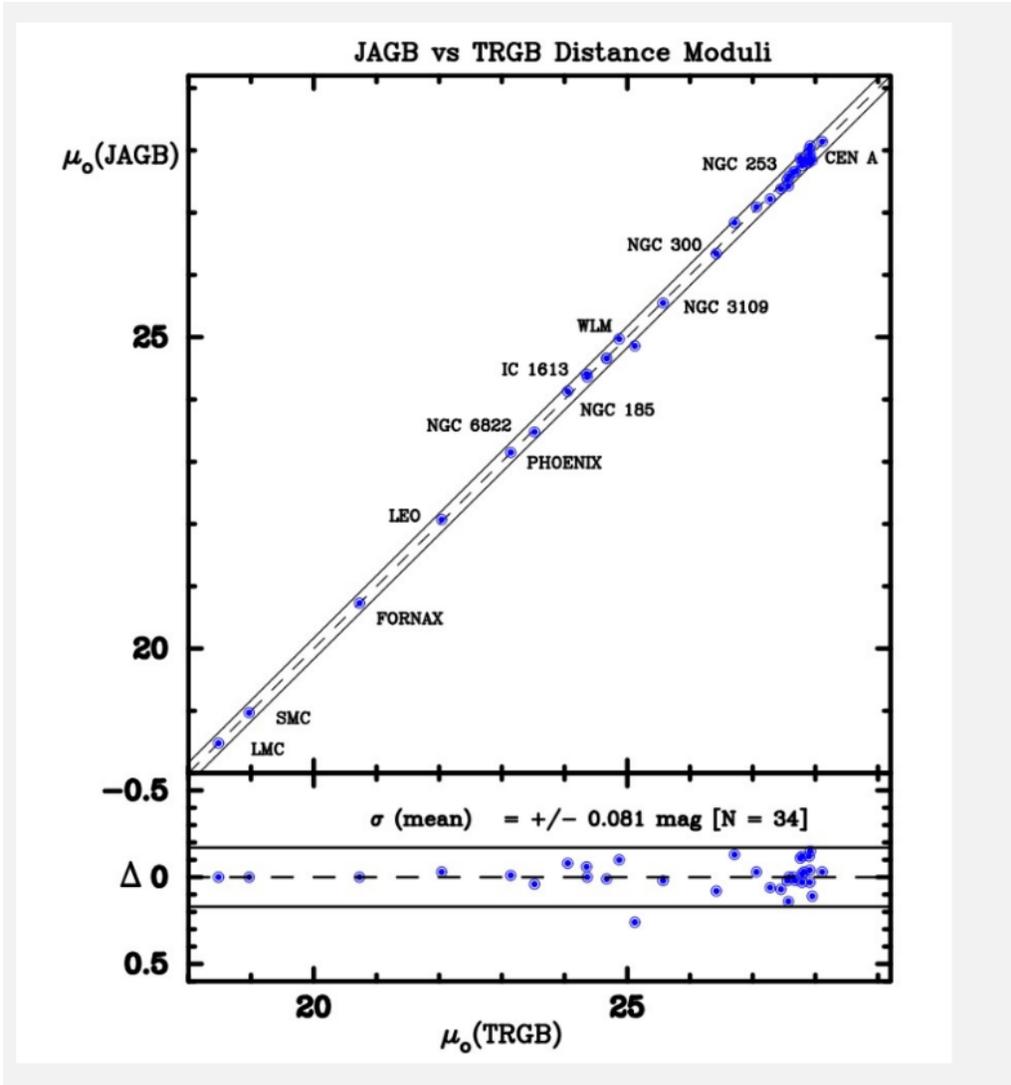

**Figure 10:** Comparison of the JAGB distance moduli of 34 nearby galaxies with respect to their independently determined TRGB distances. The inter-method scatter is only +/-0.08 mag.

# The Extragalactic (Population II) DistanceScale

With the greater light-gathering power and its optimization for observing in the infrared JWST is in a position to increase the use of the TRGB in calibrating the expansion rate of the Universe, by observing these stars in the near infrared where they are brighter than the optical and well suited to the panoramic imaging capabilities of NIRCam. An added bonus is that TRGB stars, unlike Cepheids, can be found in elliptical and S0 galaxies thereby allowing supernovae in these early-type galaxies to enter the calibration.

It was originally thought that the TRGB might resolve the tension between the Cepheid and the Planck distance scale. However, that proved not to be the case with the TRGB calibration of the Type Ia supernovae giving a value of the Hubble constant that fell midway between the competing values, as shown in Figure 11 (from Freedman 2021). And, with time, none of the three values have moved in any significant way (Figure 12).

The JAGB stars may prove to be a tie-breaker in this long-running debate. Ten galaxies that have been hosts to Type Ia supernova events are already being observed by JWST in the near and mid-infrared (JWST GO-1995, PI Freedman). Each galaxy will have Cepheid, TRGB and JAGB distances determined to them using the same instrument, thereby limiting at least one possible source of systematic error between the three methods. Progress on resolving differences between these three methods in each of the ten galaxies and in their respective calibrations of the Hubble constant will be rapid.

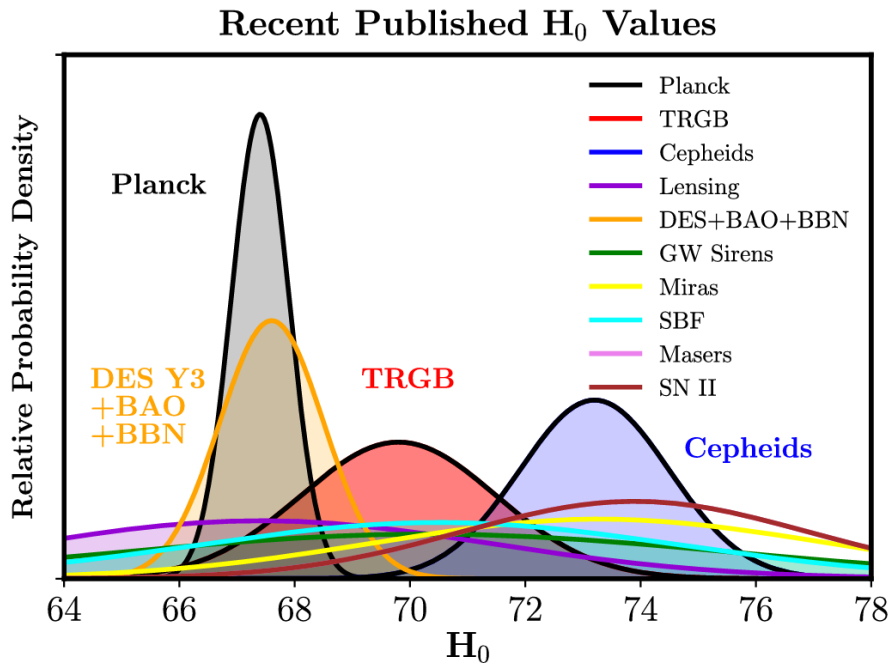

Figure 11: A graphical representation of the most recently published values of the Hubble constant based 10 different methods. Clearly, only three of the methods (Planck/CMB, TRGB and Cepheids) stand out above all of the other contenders.

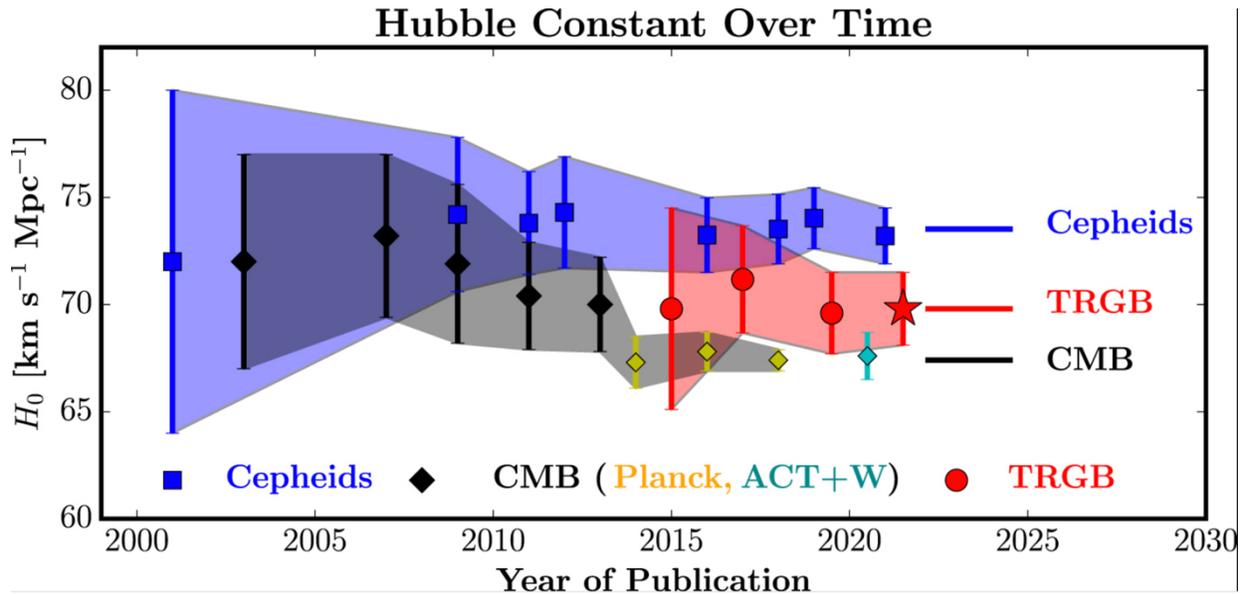

**Figure 12:** Evolution of the three most precise determinations of the Hubble Constant over time, beginning with the publication of the final results from the HST Key Project in Freedman et al. (2001). The CMB results have declined in their reported values, but within their error bars, which have decreased substantially. The Cepheid value (in blue) is still within the quoted uncertainty of the Key Project, but again, with a markedly decreased uncertainty. The TRGB value (in red) is not in meaningful contention with the Planck result (in green). Adapted from Freedman (2021).

**Conclusions**

Old and intermediate-age stellar populations now provide three independent means of gauging highly precise distances to nearby galaxies. RR Lyrae variables obey period-luminosity relations which, if properly calibrated, can provide distances to galaxies within a radius of 1-2 Mpc. In the I band, stars at the tip of the red giant branch are about 40 times brighter than the RR Lyrae variables. They are extremely well understood standard candles that have been successfully applied out to distances in excess of 20 Mpc using HST. The J-Branch AGB stars are a color-selected subset of extremely bright and very red carbon stars that are standard candles in the J band. It is expected that they can be used to determine distances out to 100 Mpc using JWST.

The tension in cosmology over the Hubble constant, as determined by the Population I Cepheids and contrasted with the cosmologically modeled value determined by the Planck mission, has been brought into question by the TRGB calibration of the Hubble constant (Freedman et al. 2020, Freedman 2021). Having all three of the methods (JAGB, TRGB and Cepheids) applied to the same galaxies, that have been hosts to Type Ia SNe, will hopefully settle this issue locally.

**Acknowledgements**

We thank our postdocs and graduate students at the University of Chicago -- Taylor Hoyt, In Sung Jang, Abigail Lee, and Kayla Owens, for their highly-valued participation in the Chicago-Carnegie Hubble Program. Financial support for this work was provided in part by NASA through grant number HST-GO-13691.003-A from the Space Telescope Science Institute, which isoperated by AURA, Inc., under NASA contract NAS 5-26555.